\documentclass{aastex}
\usepackage{epsfig}
\usepackage{emulateapj5}

\newcommand{\crj}{\underline{ \, }crj}
\newcommand{\bpic}{$\beta$~Pictoris}

\newcommand{\twhya}{TW~Hydrae}
\newcommand{\tw}{TW~Hya}

\newcommand{\kms}{km~s$^{-1}$}

\newcommand{\um}{$\mu\rm{m}$}

\newcommand{\hst}{{\it HST}}
\newcommand{\ha}{\mbox{H-$\alpha$}}

\slugcomment{Accepted for publication in ApJ on December 7, 2004}
\shorttitle{\textit{STIS} Spectroscopy and Imaging of TW~Hya}
\shortauthors{Roberge, Weinberger, \& Malumuth}

\begin{document}

%% LaTeX will automatically break titles if they run longer than
%% one line. However, you may use \\ to force a line break if
%% you desire.

\title{\textit{HST}-STIS Spatially Resolved Spectroscopy and
Coronagraphic Imaging of the TW~Hydrae Circumstellar Disk}

\author{Aki Roberge and Alycia J. Weinberger}
\affil{Department of Terrestrial Magnetism, Carnegie Institution of
Washington, Washington, DC, 20015}
\email{akir@dtm.ciw.edu, alycia@dtm.ciw.edu}

\and

\author{Eliot M. Malumuth}
\affil{Science Systems and Applications Inc., Code 681, 
Goddard Space Flight Center, Greenbelt, MD, 20771}
\email{eliot@barada.gsfc.nasa.gov}

\begin{abstract}

We present the first spatially resolved spectrum of scattered light from the 
\twhya\ protoplanetary disk.  
This nearly face-on disk is optically thick, surrounding a classical T~Tauri 
star in the nearby 10~Myr old \tw\ association.  
The spectrum was taken with the \hst-STIS CCD, providing 
resolution $R \sim 360$ over the wavelength range 5250 -- 10300~\AA.
Spatially resolved spectroscopy of circumstellar disks is difficult due to 
the high contrast ratio between the bright star and faint disk.
Our novel observations provide optical spectra of scattered light from the 
disk between 40~AU and 155~AU from the star. 
The scattered light has the same color as the star (gray scattering) at all 
radii, except the innermost region.  
This likely indicates that the scattering dust grains are larger than about 
1~\um\ all the way out to large radii. 
From the spectroscopic data, we also obtained radial profiles of the 
integrated disk brightness at two position angles, over almost the same 
region as previously observed 
in \hst-WFPC2 and NICMOS coronagraphic images 
(35~AU to 173~AU from the star).
The profiles have the same shape as the earlier ones, but show 
a small azimuthal asymmetry in the disk not previously noted. 
Our STIS broad-band coronagraphic images of \tw\ confirm the reality of 
this asymmetry, and show that the disk surface brightness interior to
140~AU has a sinusoidal dependence on azimuthal angle.
The maximum brightness occurs at a position angle of 
$233.^{\circ}6 \pm 5.^{\circ}7$ East of North.
This might be caused by the combination of forward-scattering and an 
increase in inclination in the inner region of the disk, suggesting that 
the \tw\ disk has a warp like that seen in the \bpic\ debris disk.

\end{abstract}

%% Keywords should appear after the \end{abstract} command. The uncommented
%% example has been keyed in ApJ style. See the instructions to authors
%% for the journal to which you are submitting your paper to determine
%% what keyword punctuation is appropriate.


\section{Introduction}\label{sec:intro}

\objectname[HIP 53911]{TW Hydrae} is a classical T~Tauri star, indicating that 
gas and dust from a circumstellar (CS) disk are accreting onto the young
star (spectral type K7 Ve, distance $= 56.4^{+8.1}_{-6.2}$~pc).
This disk has now been imaged at visible, near-IR, and millimeter wavelengths
(\citet{Krist:2000}, \citet{Weinberger:2002}, \citet{Wilner:2000}).
The disk inclination is $7^\circ$ from face-on \citep{Qi:2004}
and no azimuthal asymmetries have been previously reported.
A large fraction of the stellar light is reprocessed by the CS dust to 
far-IR wavelengths ($L_{\mathrm{IR}}/L_{\star} = 0.25$), indicating that the 
disk is optically thick \citep{Adams:1987}. 
These characteristics suggest that the \tw\ disk is 
relatively unevolved for its age ($\sim 10$~Myr; \citet{Webb:1999}).

However, while $\sim$1~\um\ grains in the surface layers of the disk can 
produce the mid-IR spectrum of the disk, very large (mm to cm sized) grains 
are needed to explain the mm-wavelength spectral energy distribution 
\citep{Weinberger:2002}. 
The relative lack of near-IR excess flux from the disk indicates that 
the inner 4~AU has been partially cleared of material,
possibly by formation of a giant planet \citep{Calvet:2002}.
The presence of crystalline silicates in the disk, which are seen in 
Solar System meteorites and comets but not in the interstellar medium, also 
suggests that the growth of planetary material has begun 
\citep{Weinberger:2002, Uchida:2004}.

Previous broad-band photometry indicated that the disk scattering was 
wavelength-independent and that the grain albedo might be large 
\citep{Weinberger:2002}.
This suggests that the disk is composed mostly of icy grains larger than 
$\sim$1~\um.
However, this conclusion is tentative, since it was based on broad-band 
photometry of the scattered light, rather than actual spectra. 
Also, this photometry provided no information about changes in the scattering
with location in the disk.

As part of our efforts to understand how the planetary formation process 
depends on location in the disk, we obtained the first spatially resolved 
spectrum of scattered light from a protoplanetary disk.
Our goals were to study the structure of the \tw\ dust disk, and the size 
and composition of dust grains as a function of radius from the central star.
In addition, we present broad-band coronagraphic imaging of \tw\
(Section~\ref{sec:coron}).

\section{Observations}\label{sec:obs}

\hst-STIS CCD spectra of \tw\ and a PSF star (\mbox{CD-43 2742} = HIP~32939, 
spectral type M0~V) were obtained on 2002 July~17.
All spectra were taken with the G750L grating and GAIN=1,
covering the wavelength range 5250~--~10300~\AA.
The size of the pixels in the spatial direction (y-direction) is 
0.$\arcsec$051.

The sequence of exposures was 
1) a point-source target acquisition,
2) a peak-up in a narrow slit ($52 \arcsec \times 0.\arcsec05$) to accurately 
determine the position of the star in the dispersion direction (x-direction),
3) a point-source spectrum of \tw\ taken with the 
$52 \arcsec \times 0.\arcsec2$ slit, 
4) a lamp flat field image taken with the same slit, and
5) a series of spectra of \tw\ taken with the 
$52 \arcsec \times 0.\arcsec2$~F2 slit.
This slit is the same as the slit used for the point-source spectrum,
but with a $0\arcsec.86$ wide bar placed over the central star 
(a fiducial bar).
The sequence of exposures was repeated for the PSF star.
This observing strategy might be described as ``coronagraphic
spectroscopy'', which was necessary to reduce the amount of instrumental
scattered light and allow detection of the relatively faint disk.

The resolution of the point-source spectra is 
$R = \lambda/FWHM \approx 890$ at 7000~\AA\ (corresponding
to $\sim 340$~\kms), while the resolution of the extended-source 
spectra is $R \sim 360$.
The position angle of the slit was $87.^{\circ}4$ East of North.
A log of observations appears in Table~\ref{tab:log}.

\section{Data Reduction}\label{sec:dat}

Since the $52 \arcsec \times 0.\arcsec2$~F2 slit is an unsupported observing
mode, we had to calibrate the data ourselves.  
We started with the \crj\ exposures produced by the STSDAS \emph{IRAF} 
calstis pipeline version 2.14c.
Basic data reduction had been performed on these exposures (overscan 
subtraction, bias subtraction, dark subtraction, simple flat-fielding, and 
cosmic-ray rejection). 

\subsection{Alignment of Exposures} \label{sub:align}

Before adding together the \crj\ fiducial exposures, we examined the position 
of the star in each and found that it did not vary significantly between 
exposures.
Leaving out the first \tw\ fiducial exposure for the moment 
(o64w57030\underline{ \, }crj.fits), the mean x and y shifts between exposures 
were 0.007 pixels and 0.003 pixels.
The first \tw\ fiducial exposure was short (only about 14\% of the exposure 
time of the others) and had much lower signal-to-noise, making determination 
of the star position less precise.  
Still, the maximum x and y shifts of the first exposure relative to 
any other were still very small (0.04 pixels in x and 0.09 pixels in y).
Therefore, the five \tw\ \crj\ fiducial exposures were added together with 
no realignment of the images and the exposure time in the file 
header changed to the sum of the individual exposure times.
The shifts in position of the star between the two PSF fiducial exposures were 
very small (0.002 pixels in x and 0.005 pixels in y), so they were also 
combined with no realignment.

\subsection{Defringing \& Calibration}\label{sub:defr}

\begin{figure}
\epsscale{1.0}
\plotone{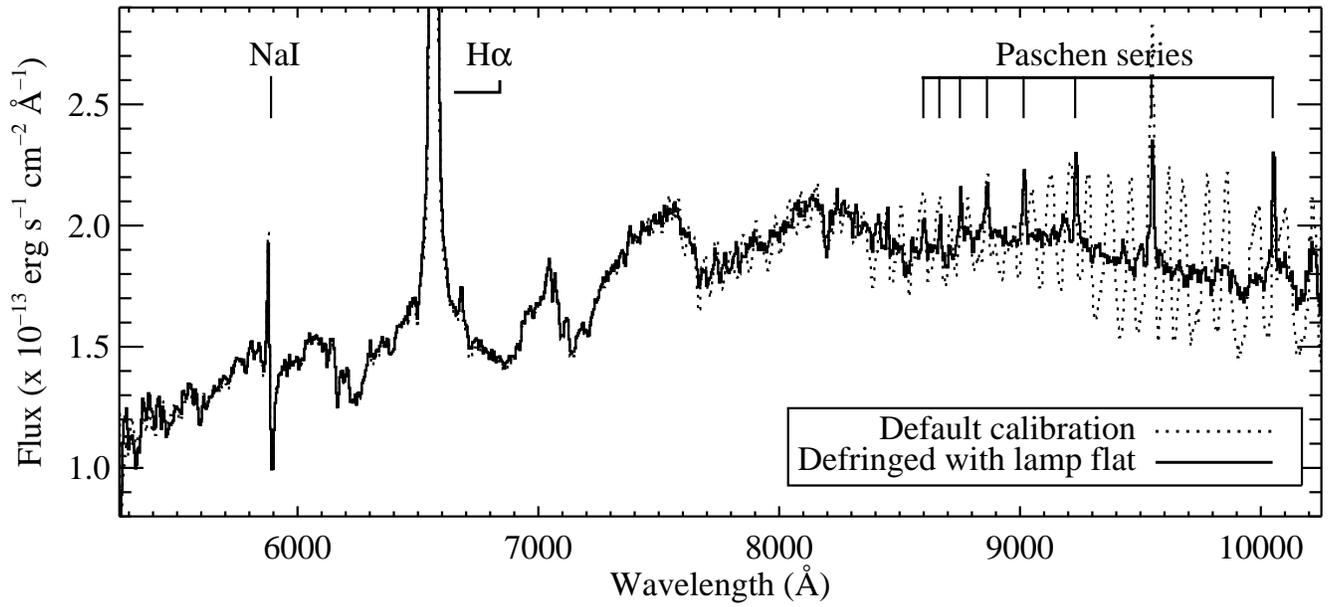}
\caption{Point-source STIS G750L spectrum of \tw. The default calibrated
data are shown with a dashed line.
The defringed data are shown with a solid line.
Defringing allows detection of several Paschen series \ion{H}{1} 
emission lines;
these lines are caused by accretion of disk gas onto the star, as are
the \ion{Na}{1} and \ha\ emission lines.
\label{fig:pt}}
\end{figure}

STIS CCD spectra suffer from fringing at long wavelengths, caused by 
interference between reflections from the front and back surfaces
of the CCD (see Figure~\ref{fig:pt}).
Using our contemporaneous lamp flat images and \emph{IRAF} tasks 
provided by STSDAS (prepspec, normspflat, mkfringeflat, and defringe), 
the fringing was removed from the point-source and fiducial data of 
both \tw\ and the PSF star.
For the fiducial data, the defringing was performed on the combined
images, since the higher signal-to-noise allows better defringing than is 
possible on the separate exposures.

After defringing, we were left with combined 2-D spectra in counts per pixel.  
These were wavelength calibrated using the STSDAS \emph{IRAF} task
wavecal. 
The point-source data were flux calibrated using the STSDAS task 
x1d, which extracts one-dimensional spectra from the images. 
The fiducial data were flux calibrated with the STSDAS task x2d, producing
pixels with units of \mbox{erg s$^{-1}$ cm$^{-2}$ \AA$^{-1}$ arcsec$^{-2}$}.
This task also applies a geometrical distortion correction, producing 2-D 
spectra with wavelength running linearly along the x-axes of the images and
distance along the slit running linearly along the y-axes.

\subsection{Correction of Hot and Cold Pixels} \label{sub:clip}

We found that the best weekly average dark image used by the calstis pipeline 
still left a number of hot and cold pixels in the data.
We therefore performed a sigma-clipping procedure tailored for 2-D spectra on 
our calibrated fiducial images.  
The procedure was tailored primarily by using a clip box that mimicked 
the shape of the data, i.e.\ 1201 pixels in x (full x-range of the image) by 
5 pixels in y.
This box worked best since the values along a row of a 2-D spectrum don't 
vary as much as they do in the y-direction; a standard square clip box 
tended to miss hot and cold pixels in the portion of the image containing 
the spectrum.

If the brightest (and faintest) pixel in each clip box varied from the median 
value of its neighboring pixels by more than 3 times the local noise,
it was replaced by the median value.
The local noise was the standard deviation of the values in the clip box
(spectrum noise) added in quadrature to the standard deviation of the
column containing the brightest or faintest pixel (background noise).
The background noise was determined in this way since it had a clear 
dependence on x-position (the background is worse on the long-wavelength
end of the CCD). 
This process was applied to the data until the number of deviating pixels 
approached zero.  
This clipping procedure did not correct unreasonably large numbers of pixels
(only $0.16\%$), remove background noise, 
or preferentially treat the portion of the image containing the spectrum.

\begin{figure}
\plotone{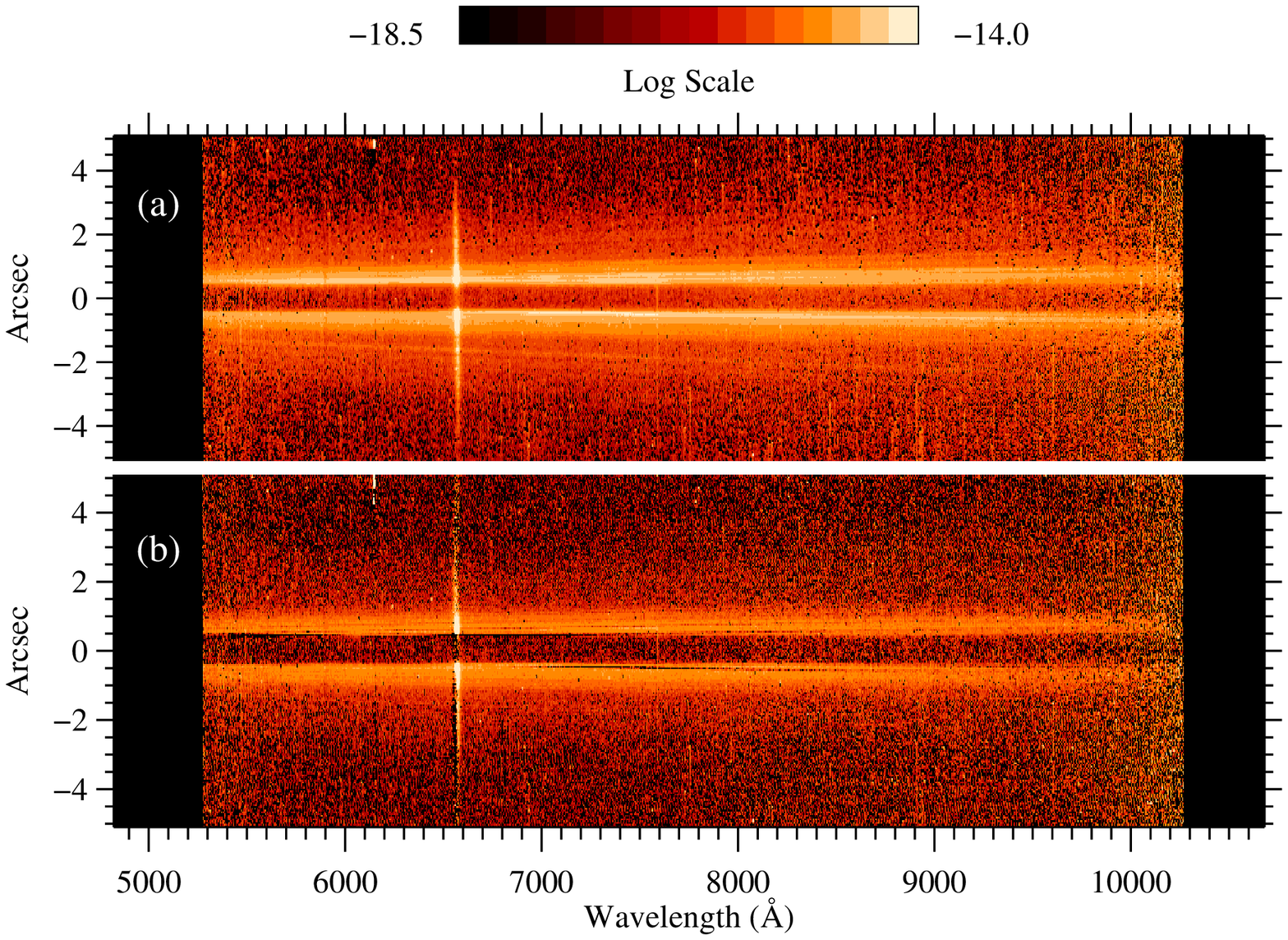}
\caption{STIS G750L fiducial spectra of \tw.
The dispersion direction is along the x-axis and the spatial direction
along the y-axis.
The figure shows a logarithmic scaling of surface brightness in units
of \mbox{erg s$^{-1}$ cm$^{-2}$ \AA$^{-1}$ arcsec$^{-2}$}.
The central star is blocked by the $0.\arcsec86$ wide fiducial bar.
\ \textit{(a, top)} \ The total combined \tw\ 2-D spectrum.  
The stellar \ha\ emission line is visible at 6560~\AA.
Linear streaks caused by dispersal of the Airy rings of the telescope
PSF can been seen angling away from the fiducial bar from left to right
across the image.
\ \textit{(b, bottom)} \ The PSF-subtracted 2-D spectrum of \tw.
The light seen is scattered by the CS dust disk.
The surface brightness at \ha\ is somewhat overestimated; see 
Sections~\ref{sub:psf} and \ref{sub:RI_prof} for discussion of this issue.
\label{fig:2d}}
\end{figure}

The final calibrated 2-D \tw\ fiducial spectrum appears in the top panel of 
Figure~\ref{fig:2d}.
The primary instrumental features of the 2-D spectrum are a smooth halo of 
scattered light and linear streaks caused by dispersing the Airy rings of 
the telescope PSF.  
These streaks angle away from the central star from left to right across 
the image, since the size of an Airy ring increases with wavelength
(see \textit{STIS Instrument Handbook, version 6.0}, Section 13.7.3 for
an explanation).

\subsection{PSF Subtraction} \label{sub:psf}

The next step in the reduction of the \tw\ data was subtraction of the stellar 
light from the fiducial image in order to isolate light reflected from the CS 
disk.  
The PSF star was chosen to match the spectral type of \tw\ as closely as 
possible, but it is not an exact match.  
To correct for the difference in overall brightness and spectral mismatch 
between \tw\ and the PSF star, we made use of the point-source spectra of 
both stars.  
Each column of the PSF fiducial image was scaled by the ratio of the \tw\ and 
PSF star point-source spectra at that wavelength, to produce a synthetic 
PSF image with the same brightness and color as \tw. 
The statistical errors of the point-source spectra were propagated into the 
error array of the synthetic PSF image. 

The synthetic PSF image was then subtracted from the \tw\ fiducial image; the
result is shown in the bottom panel of Figure~\ref{fig:2d}. 
The remaining continuum flux seen is light reflected from the CS dust disk.
At $1 \arcsec$ from the star, the surface brightness of the disk is slightly 
greater than the surface brightness of the light coming from the star 
($S_{\textrm{disk}} = 1.16 \times S_{\textrm{star}}$), demonstrating the 
excellent suppression of scattered stellar light by the fiducial bar.

The disk surface brightness at \ha\ is somewhat greater than it should be, 
since the emission line in the \tw\ point-source spectrum was saturated, 
while the line in the fiducial exposures was not. 
This resulted in underestimated PSF scaling ratios at the saturated 
wavelengths. 
We were not able to correct the saturated pixels since the nonlinearity
beyond saturation of the STIS CCD at GAIN=1 has not been characterized
\citep{Gilliland:1999}.
The large scale factor at \ha\ (caused by the fact that the PSF star doesn't 
show the emission line) also produced a stripe of increased noise in the 
PSF-subtracted image.
The \ha\ wavelength region will be ignored in the disk spectra 
discussed in Section~\ref{sec:spec}.
A further discussion of this problem appears in Section~\ref{sub:RI_prof}.

\section{Sources of Systematic Error} \label{sec:error}

The three major factors limiting the quality of a PSF subtraction are 
1) a change in the shape of the telescope point-spread function, 
2) an offset in position between the target star and the PSF star, and 
3) a color mismatch between the target star and the PSF star. 
The first factor is caused primarily by thermal flexure of the telescope; 
there is nothing that can be done post-observation to correct this.  
The only way to minimize the effect is to choose a PSF star that is near
the target star on the sky, so that the telescope is oriented in the same
way relative to the Sun, and to observe the target and PSF stars close 
together in time, which we did (observed in consecutive orbits).
In Sections~\ref{sec:prof} and \ref{sec:spec}, we will discuss the artifacts 
introduced by this effect. 
Here we discuss our pre- and post-observation efforts to minimize the 
other two effects and characterize the systematic errors introduced.

\subsection{Misalignment of \tw\ and the PSF} \label{sub:error_align}

An offset between the positions of \tw\ and the PSF star in the fiducial 
images will obviously result in a degradation of the quality of the PSF 
subtraction.  
To ensure that any offset was as small as possible, a peak-up along the 
dispersion direction was performed after acquisition of each star.
This procedure centers the star in the science slit with an accuracy of 5\% 
of the peak-up slit width, which in our case corresponds to 0.049 x-pixels 
\citep{Brown:2002}.
Such a small offset does not significantly affect the quality of our PSF 
subtraction.
However, we did not perform a peak-up along the spatial dimension to center 
the star in the y-direction.
A STIS point-source target acquisition typically centers the star to within 
$0.\arcsec01 = 0.2$ pixels \citep{Brown:2002}.
An offset between \tw\ and the PSF star of a few tenths of a pixel would 
introduce significant systematic error.

We attempted to measure the offset between the positions of \tw\ and the PSF 
star in the fiducial spectra several ways.
Unfortunately, alignment of the two images appears to be dominated by small 
scale structure, and different techniques gave different offsets.  
As will be seen in Section~\ref{sec:prof}, an offset of 0.0 y-pixels produces 
a PSF-subtracted 2-D spectrum with similar total disk fluxes above 
and below the fiducial.
Since we do not expect the disk to be much brighter on one side than the 
other, we chose our best y-offset to be zero. 
In order to investigate the effect of a small misalignment in the y-direction, 
we produced PSF-subtracted images in which the PSF star image was shifted 
relative to the \tw\ fiducial image by $\pm 0.25$ y-pixels.  
The changes seen in the radial surface brightness profiles and the extracted 
disk spectra (discussed in Sections~\ref{sec:prof} and \ref{sec:spec}) 
were used to characterize the systematic error introduced by a misalignment 
of the \tw\ and PSF stars.  

\subsection{Incorrect Scaling of the PSF Star to \tw} \label{sub:error_color}

Color mismatch between the target star and the PSF star can be a major 
source of systematic error during PSF subtraction, since the size of 
the telescope PSF is a function of wavelength. 
In the case of \tw, a mismatch in total brightness may also occur, since this 
star is variable in brightness and color on various timescales.
The Cousins R and I band variability is shown below \citep{Mekkaden:1998}. 
\vspace{1ex}
\[ \begin{array}{l@{\extracolsep{1ex}}l@{\extracolsep{1ex}}r@{\extracolsep{2ex}}l@{\extracolsep{1.5ex}}c@{\extracolsep{1.5ex}}r}
\mathrm{Over} & 1.13 & \mathrm{years} : & 
\mathrm{R}_{\textrm{mean}} \pm 1 \sigma & = & 10.101 \pm 0.061 \\
 &  &  &  
 \mathrm{I}_{\textrm{mean}} \pm 1 \sigma & = & 9.341 \pm 0.031 \\
\mathrm{Over} & 8.03 & \mathrm{days} : & 
\mathrm{R}_{\textrm{mean}} \pm 1 \sigma & = & 10.126 \pm 0.048 \\
 &  &  &  
 \mathrm{I}_{\textrm{mean}} \pm 1 \sigma & = & 9.351 \pm 0.028 \\
\mathrm{Over} & 4.12 & \mathrm{hours} : & 
\mathrm{R}_{\textrm{mean}} \pm 1 \sigma & = & 10.128 \pm 0.013  \\
&  &  &  
 \mathrm{I}_{\textrm{mean}} \pm 1 \sigma & = & 9.351 \pm 0.009 \vspace*{1ex}
\end{array} \]  

We addressed the problems of variability and color mismatch by obtaining 
point-source spectra of both stars before the fiducial observations, so 
that we would be able do a near-time, wavelength-dependent scaling of the 
PSF star to \tw. 
The expected R band variability of \tw\ over the time between the point-source 
observation and the end of the fiducial observations (approximately 2 HST 
orbits $\approx$ 2 hours) is $\pm 0.0063$, corresponding to a $\pm 0.58$\% 
flux variability; the I band variability is smaller.
The star should have been stable to within the uncertainty of the STIS flux 
calibration. 
We confirmed this by examining the brightness of \tw\ in the five fiducial 
exposures, and found that it was stable at all wavelengths at the 1\% level. 

Even if the brightness and color of \tw\ were stable over the course of our 
observations, there still might be an error in the absolute flux calibration 
of \tw\ or the PSF star. 
We examined this by adjusting the point-source fluxes by $\pm 1$\%,
producing altered ratio arrays and mis-scaled synthetic images.
This changes the overall brightness of the synthetic PSF images and 
introduces a small color mismatch as well, since the PSF star does not
have a flat spectrum.
Generally, this introduced a systematic error of $\pm 2 \%$ to $\pm 4 \%$ 
in the disk radial profiles and the extracted disk spectra.
A few specific pixels in the outer regions of the disk where the signal is 
low showed larger percent errors.  
But as a whole, the error introduced by a reasonable estimate of the possible
mismatch between the scaled PSF image and the \tw\ fiducial image is 
consistently smaller than the statistical uncertainty of the data.

\section{Disk Radial Profiles} \label{sec:prof}

\subsection{Total Profiles} \label{sub:tot_prof}

\begin{figure}
\plotone{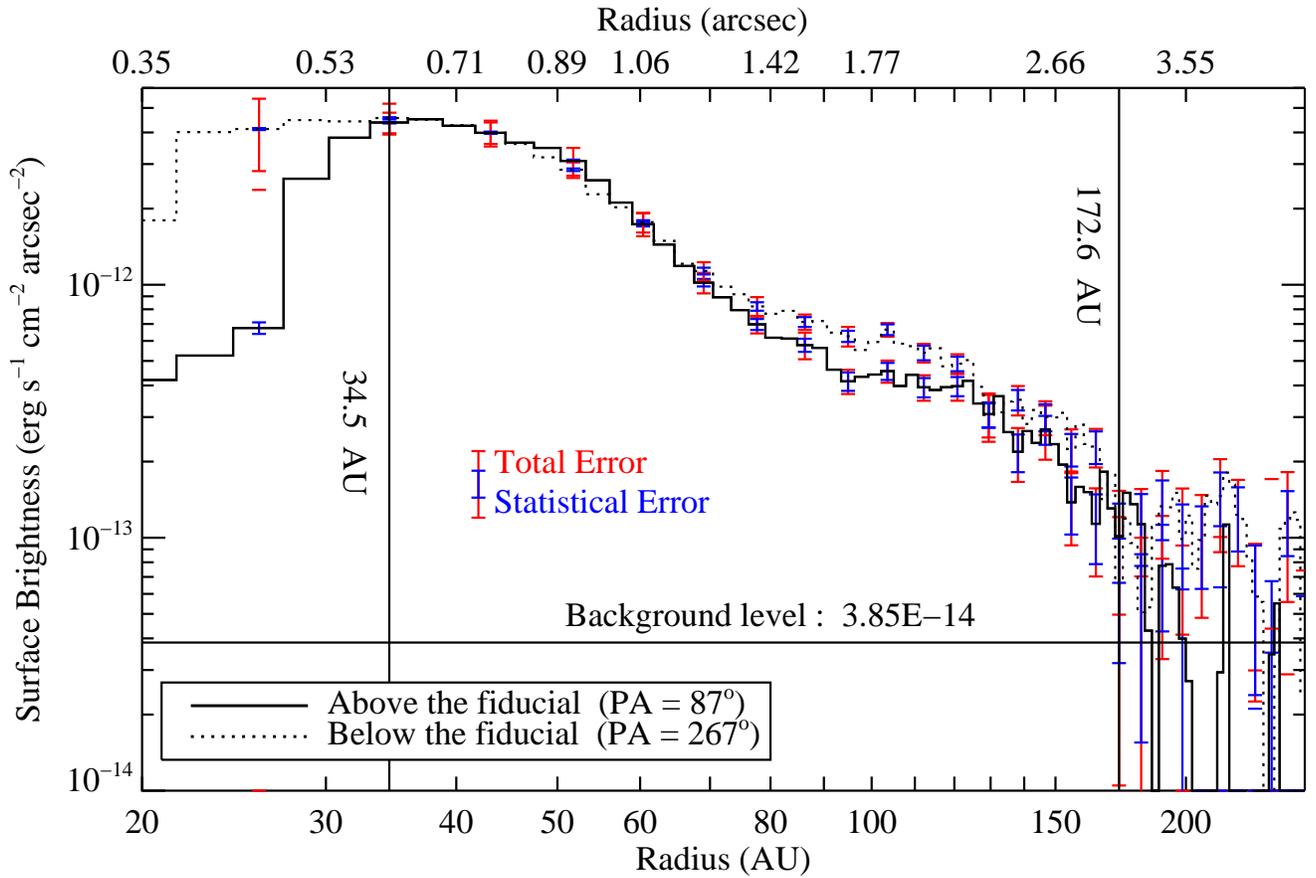}
\caption{Radial surface brightness profiles of the \tw\ disk, integrated
over the whole G750L bandpass ($5250 - 10300$~\AA).
The profile above the fiducial bar (position angle = $87.^{\circ}4$ E of N)
is shown with a solid line, the profile below (PA = $267.^{\circ}4$ E of N) 
with a dashed line.
The statistical errors are shown with blue error bars; 
the total errors, including systematic errors, are shown with red bars.
The disk is detected between 34.5~AU and 172.6~AU.
The background level (the median surface brightness outside of 200~AU)
is indicated with a horizontal line.
\label{fig:tot_prof}}
\end{figure}

Total disk radial surface brightness profiles were produced by integrating 
every row of the 2-D PSF-subtracted disk spectrum over the whole G750L 
wavelength band (including the \ha\ emission line).
The profiles above and below the fiducial bar are shown in 
Figure~\ref{fig:tot_prof}.
Note that ``above the fiducial bar'' corresponds to a disk position angle
of $87.^{\circ}4$ east of north and ``below the fiducial bar'' corresponds to
PA = $267.^{\circ}4$.
The blue error bars are the propagated statistical errors of the data, 
while the red error bars include the systematic error introduced by 
misalignment of \tw\ and the PSF image (as discussed in 
Section~\ref{sub:error_align}).
The statistical errors exceed the systematic errors outside of 
$1\arcsec.02 = 57.5$~AU.
The inner radius at which we claim detection of the disk (34.5~AU) is the 
innermost radius at which the profile brightnesses above and below the 
fiducial are within one sigma of each other. 
The outermost radius (172.6~AU) is the point at which the profiles begin 
to be within one sigma of the background level (the median surface 
brightness outside of 200~AU).

The total disk fluxes are similar above and below the fiducial: 
$F_{\mathrm{above}} = (6.07 \pm 0.17) \times 10^{-13}$ erg s$^{-1}$ cm$^{-2}$
and 
$F_{\mathrm{below}} = (6.44 \pm 0.16) \times 10^{-13}$ erg s$^{-1}$ cm$^{-2}$.
However, the disk is noticeably brighter below the fiducial between
$r = 78$~AU and $r = 124$~AU.
The difference in surface brightness between the two profiles, integrated 
over this range of radii, is significant at the $8 \sigma$ level.
The peak difference in surface brightness occurs at 95~AU.
This azimuthal asymmetry, the first seen in the \tw\ disk, is 1$\arcsec.38$ 
to 2$\arcsec.19$ from the star, where the statistical uncertainties dominate 
over the systematic uncertainties and there are no visible PSF-subtraction
artifacts.  
In Section~\ref{sec:coron}, we will discuss our STIS broad-band coronagraphic 
imaging of the disk, which also shows azimuthal asymmetry in the disk
brightness.

\subsection{R and I Band Profiles} \label{sub:RI_prof}

We also produced Cousins R and I band radial surface brightness profiles for 
comparison to previously published \hst\ WFPC2 profiles.
The conversion to surface brightness in magnitudes arcsec$^{-2}$ was done 
by applying the following equation to every row of our 2-D disk spectrum:

\begin{equation} \label{eq:ri}
S^{'}_{\mathrm{disk}} = \mathrm{m}_{\mathrm{Vega}} 
- 2.5 \times \log{  \frac{ \int_{0}^{\infty} \ q(\lambda) \, S_{\mathrm{disk}}%
(\lambda) \, \mathrm{d} \lambda}{%
\int_{0}^{\infty} \ q(\lambda) \, F_{\mathrm{Vega}}(\lambda) \, \mathrm{d}% 
\lambda}},
\end{equation}
where $S^{'}_{\mathrm{disk}}$ is the disk surface brightness in magnitudes 
arcsec$^{-2}$, $\mathrm{m}_{\mathrm{Vega}}$ is the magnitude of Vega (0.0), 
$q(\lambda)$ is the Cousins filter throughput curve from the \emph{IRAF} 
SYNPHOT package input tables, $S_{\mathrm{disk}}(\lambda)$ is the disk surface 
brightness in units of \mbox{erg s$^{-1}$ cm$^{-2}$ \AA$^{-1}$ arcsec$^{-2}$} 
(the pixel values in our 2-D disk spectrum), and $F_{\mathrm{Vega}}$ is the
spectrum of Vega from SYNPHOT in units of 
\mbox{erg s$^{-1}$ cm$^{-2}$ \AA$^{-1}$}.
The resulting R and I band radial profiles are shown in Figure~\ref{fig:RI}.
The asymmetry seen in the total profile in Figure~\ref{fig:tot_prof} is seen 
in both the R and I band profiles.

The WFPC2 profiles overplotted in Figure~\ref{fig:RI} are the 
profiles shown in Figure~5 of \citet{Krist:2000}, converted from 
WFPC2 F606W and F814W band fluxes per arcsec$^{2}$ to R and I band 
magnitudes per arcsec$^2$.
Our profiles show the same decrease in slope between 79 and 134~AU 
labeled Zone~3 by \citet{Krist:2000} and also seen in \hst\ NICMOS
profiles of the disk \citep{Weinberger:2002}.
We note that the asymmetry between the profiles above and below the 
fiducial occurs over almost exactly the same region.

There is extraordinarily good agreement between the absolute brightnesses 
of our profiles and WFPC2 profiles, which may be somewhat surprising 
since the star is variable.
We calculated the R and I band magnitudes of the star at the time 
of our observations by replacing $S_{\mathrm{disk}}(\lambda)$ in 
Equation~\ref{eq:ri} with the point-source \tw\ spectrum.
The magnitudes of \tw\ were R~=~$10.1875 \pm 0.0083$ (including the
\ha\ emission line) and I~=~$9.4611 \pm 0.0064$, slightly fainter than
the mean long-term magnitudes shown in Section~\ref{sub:error_color}.
The $\pm 1 \sigma$ errors take into account the 5\% absolute photometric 
accuracy of the STIS flux calibration \citep{Brown:2002}.

Saturation of the \ha\ emission line in the \tw\ point-source spectrum 
will cause our stellar R band magnitude to be somewhat underestimated.
We examined this problem by comparing the observed equivalent width of 
the line to the \ha\ widths seen during a 2~year-long \tw\ spectroscopic 
monitoring project \citep{Alencar:2002}.
We measured an \ha\ equivalent width of $280.5 \pm 6.5$~\AA, which is a
lower limit to the true equivalent width.  
This value is very close to the largest equivalent width seen 
(about $274 \pm 19$~\AA; \citet{Alencar:2002}).
This shows that although the photosphere of \tw\ was fainter than usual 
during our observations, the \ha\ line was quite bright; 
this is not strange, since there is no correlation between the \ha\ line
strength and the stellar photospheric flux \citep{Mekkaden:1998,Alencar:2002}.
The large equivalent width of our line also indicates that the \ha\ 
point-source flux and stellar R band magnitude are probably not greatly 
underestimated.

The R and I band magnitudes of \tw\ during the WFPC2 observations in
\citet{Krist:2000} were R~=~$9.84 \pm 0.06$ and I~=~$9.16 \pm 0.06$,
which are 0.35 and 0.30 magnitudes brighter than the 
star at the time of our observations.
The expected 0.35 and 0.30 magnitude arcsec$^{-2}$ differences between our 
R and I band profiles and the WFPC2 profiles are not seen.
We do not know the reason for the close agreement between the profiles. 

\section{Disk Spectra} \label{sec:spec}

One-dimensional spectra at increasing radii from the star were 
extracted from the PSF-subtracted 2-D \tw\ disk spectrum.
The surface brightness values were integrated over the spatial dimension
of the extraction box, then converted to flux units using the DIFF2PT 
header keyword produced by the STSDAS x2d tool.
This keyword uses a wavelength-averaged aperture throughput value;
the extracted spectra have the flux values that are correct in the
center of the bandpass, but are too high and too low at the ends.
This effect is removed when we divide our extracted disk spectra by the 
point-source stellar spectrum extracted in the same way.

The height of the extraction box was chosen to be 5~pixels 
($0.\arcsec255$). 
Spectra extracted using shorter boxes showed peaks and dips due to PSF 
structure not completely removed by the PSF-subtraction technique.
More severe PSF-subtraction residuals near the fiducial bar are the 
limiting factor on the radius of the innermost extraction box;
we avoided the 4~pixels just above and below the fiducial. 
This centers the innermost spectrum at 40.3~AU, including radii 
from 34.5~AU ($0.\arcsec612$) to 46.0~AU ($0.\arcsec816$). 
Avoiding the 4~pixels nearest the fiducial bar places the innermost pixel 
of the extraction box at the innermost radius of the disk profiles 
shown in Figure~\ref{fig:tot_prof}.
The PSF-subtraction residuals and their effect on the extracted spectra
will be discussed in more detail later in this section.

The disk spectra have the same shape at all radii above and below the 
fiducial bar, so they were averaged together to increase the signal-to-noise.
The extracted disk spectra divided by arbitrarily scaled 
\tw\ point-source spectrum (the disk color spectra) are shown in 
Figure~\ref{fig:color}, rebinned by a factor of 10 to increase the
$S/N$.
Misalignment of \tw\ and the PSF star (discussed in 
Section~\ref{sub:error_align}) does not significantly affect the 
disk color spectra.
The error bars plotted in Figure~\ref{fig:color} are the propagated
statistical errors.
The $S/N$ per resolution element (2~x-pixels) of the spectra are
15 at 40.3~AU, 13 at 54.7~AU, 7 at 69.0~AU, 4 at 97.8~AU, and 1 at 155.3~AU.
The disk color spectra are flat between 55~AU and 155~AU, indicating
that the scattering function is wavelength-independent over the bulk of 
the disk (neutral or gray scattering).

In the innermost spectrum, the disk-to-star ratio increases at shorter 
wavelengths; the disk scattering appears significantly bluer than it
does at larger radii.
Although we have avoided the worst residuals near the fiducial, we must
carefully consider whether this feature is real or due to imperfect PSF
subtraction.
Unfortunately, the data do not exist to accurately quantify the effect of PSF 
subtraction residuals in STIS fiducial spectra, since very few observations 
were taken in this mode. 
We tried to quantify the effect of residuals in our data in several ways, 
including analysis of a set of G750L spectra taken 
with $52\arcsec \times 0\farcs2$~F2 under calibration program 8844.
But these observations were not planned with PSF stability in mind; 
the best subtraction we could achieve, using the calibration observations 
closest in time, of the two stars most closely matched in spectral type, was 
obviously much worse than what we achieved for \tw. 

We are reduced to qualitative arguments about the reality of the blue 
scattering seen in the spectrum extracted at 40.3~AU. 
In Figure~\ref{fig:artifact}, the innermost region of the PSF-subtracted
2-D disk spectrum is shown at the bottom. 
The worst PSF-subtraction residual is visible as a dark stripe just
below the fiducial bar.
The spectra extracted at 28.8~AU and 40.3~AU are shown at the top of
the figure;
the white bars at the edge of the disk image show the limits of the
extraction boxes in the y-direction.
The spectrum at 28.8~AU is just at the edge of the fiducial bar, and 
includes the worst residual.
It may be seen in the extracted spectrum as a roughly 30\% dip between 
about 7000~\AA\ and 8400~\AA.

Since the PSF-subtraction residuals angle away from the fiducial bar from
left to right, it might be possible that the worst residual moves into
the 40.3~AU extraction box at longer wavelengths.
However, the depth of the decrease in the disk color spectrum at long 
wavelengths is about 30\%.
A residual that could make such a deep dip in the extracted spectrum 
should be visible in the 2-D spectrum; there is no visible residual  
in the 40.3~AU extraction box.
Also, the disk color spectrum extracted above the fiducial  
is the same as that extracted below, i.e.\ bluish.
There is no obvious PSF-subtraction residual above the fiducial that 
could cause this.
The blue scattering in the 40.3~AU spectrum is probably real,
but we are not able to prove it conclusively.

\section{STIS Coronagraphic Imaging of TW~Hya}\label{sec:coron}

\subsection{Observations}

STIS CCD images of \tw\ and a different PSF star (HD~85512), 
were obtained on 2000 November 3 and 2000 December 27. 
After a standard point-source target acquisition, the
stars were observed with the WEDGEA1.0 coronagraphic aperture.  
This is a 1\arcsec\ wide location on the tapered wedge parallel to the 
CCD charge transfer direction.  
The STIS CCD is unfiltered, covering essentially the entire visible bandpass. 
The stars were observed with GAIN=4 to increase the dynamic range of the 
images. 
At this gain, the electronics read out the full well of the CCD. 
Individual frames of 226~s were acquired on \tw\ and 56~s on the PSF, 
chosen to be longer than the read-out time of the CCD for increased 
efficiency and short enough to avoid saturating the pixels just outside the 
wedge. 
The \tw\ field rotated 33$^\circ$ with respect to the STIS CCD 
between the two visits.
The coronagraphic imaging observations are listed in Table~\ref{tab:log}.

\subsection{Data Reduction}

Calstis performs all of the necessary image reduction, including
combination of the CR-SPLIT exposures. 
We then combined the repeated exposures, weighting by their integration 
times, to produce an image in counts per sec.
Although bias is supposed to be subtracted in this process, we found that the 
CCD frames were distinctly non-zero in the portions well away from the disk.  
We took the median value of the pixels in the field and subtracted it from the
frame to bring the bias to zero.

\subsection{PSF Subtraction}

As with the spectroscopy, to remove the stellar point spread function
from the image of the disk, a PSF star must be subtracted. 
We used the
diffraction spikes to align and scale the PSF star separately 
to the two
visits of \tw, finding the best subtraction through $\chi^2$
minimization. 
This process reveals a (-0.53, 0.00) pixel shift between
the PSF and \tw\ in the first visit and a (-0.08, -0.72) shift for the
second visit. 
The $\pm 1 \sigma$ error in the alignment is $\pm 0.027$ pixels in x and y.
The best scale factors indicate that the PSF was 
3.16~mag brighter than \tw\ in November and 3.08~mag brighter than \tw\ in
December. 
The SIMBAD database gives the V magnitude difference 
between the two stars as 3.2 mag, so these scales are reasonable.
Assuming that the PSF does not vary in brightness, these scalings show
that \tw\ was 0.08 mag brighter in November than December. This is
within the $2 \sigma$ variability of \tw\ in the V and R-bands on month
time-scales \citep{Mekkaden:1998}.

Next, the December PSF-subtracted image was scaled and registered to the
November PSF-subtracted image using the offsets and scale factors
measured above.  Both images were rotated to put North up and East
left. Then, the wedges and four pixel wide swaths across the diffraction 
spikes were masked.  The two images were combined by averaging pixels that 
were unmasked in both images or using pixels that were unmasked in one or
the other of the images. The final image is shown in Figure~\ref{fig:coron}. 
Pixels that were masked in both orientations are blacked out in this image.
The slit used in the spectroscopic observations is overlaid in white
for illustration.

\subsection{Results}
\label{sub:coron_results}

The radial surface brightness profiles from the coronagraphic image are
shown in Figure~\ref{fig:coron_prof}.  
The azimuthally-averaged radial profile shows that the disk is detected from 
33~AU ($0.\arcsec59$) to 283~AU ($5.\arcsec02$) in the final coronagraphic 
image.  
The inner radius is set by the size of the coronagraphic wedge and the outer
radius is the point outside of which the surface brightness is less than
3 times its uncertainty.

Although the disk looks quite circular by visual inspection, a
quantitative analysis demonstrates that it is asymmetric.
Figure~\ref{fig:best_asym} shows the brightness of the disk between 70 and 
88~AU as a function of position angle in $20^{\circ}$ intervals.
The disk brightness peaks at PA = $233.^{\circ}6 \pm 4.^{\circ}0$ and is 
well fit by a sine function;
this error bar is the $\pm 1 \sigma$ statistical error.  
The direction of maximum brightness is indicated with a green
line on the disk image in Figure~\ref{fig:coron}.
The ratio of the disk brightnesses in the directions of minimum and
maximum brightness is $0.52 \pm 0.04$ (again, $\pm 1 \sigma$ statistical
error).
The min/max brightness ratio is relevant to analysis in 
Section~\ref{sub:asym} below.

To estimate the systematic errors introduced by misalignment of the
PSF and \tw, we offset the PSF image by $\pm 3 \sigma = 0.08$~pixels 
along the direction of maximum brightness.
Offsets in this direction should produce the largest changes to the 
azimuthal brightness asymmetry.
The offset PSF image was subtracted from \tw\ and the sine fit to the
brightness vs.\ position angle recalculated.
We find that the change to the min/max brightness ratio is only
$0.02$, which is smaller than the statistical uncertainty ($0.04$).
To estimate the total systematic error in our PSF subtraction, 
we compared the results from the sine fit to the combined data to 
the results from fits to the November and December data separately.
The difference in the maximum brightness direction between the two visits 
was $4^{\circ}$, the same as the statistical uncertainty of $4^{\circ}$ 
quoted above. 
Our conservative estimate of the systematic error in the min/max
brightness ratio is $\pm 0.15$. 
This shows that most of the systematic error is not due to misalignment,
but more likely due to changes in the shape of the PSF

The profiles including only pixels with position angles within 
$\pm 20^{\circ}$ of the directions of maximum and minimum brightness are also
shown in Figure~\ref{fig:coron_prof}.  
Unfortunately, the maximum brightness profile (PA~$=233.^{\circ}6$) only 
reaches in to 65~AU and the minimum brightness profile (PA~$= 53.^{\circ}6$) 
only reaches in to 62~AU, due to the position of the coronagraphic wedge in 
both visits.
But the profiles clearly show that the disk is significantly brighter on
one side between 65~AU and 140~AU.  
The difference in surface brightness between the two profiles, summed over 
this range of radii, is significant at the $10 \sigma$ level.  
Exterior to 140~AU, the disk is symmetric.

The position angle of the slit in the spectroscopic observations ``below
the fiducial'' is only $34^{\circ}$ away from the direction of maximum
brightness, and comparison of the profiles in Figures~\ref{fig:tot_prof}
and Figure~\ref{fig:coron_prof} is instructive.  The profiles are
qualitatively similar, both showing that the disk is brighter in the
``below the fiducial'' direction over some range of radii around $r \sim
100$~AU.  However, the coronagraphic profiles show that the disk is
significantly asymmetric over a larger range of radii than the
spectroscopic profiles do.  The reason for this may be the larger error
bars of the spectroscopic profiles and/or the arbitrary alignment of the
\tw\ and synthetic PSF fiducial images in the y-direction before
subtraction, as discussed in Section~\ref{sub:error_align}.

\section{Discussion}\label{sec:disc}

\subsection{Azimuthal Asymmetry in the Disk Radial Profiles} \label{sub:asym}

Our spectroscopic and coronagraphic disk radial surface brightness
profiles show the first azimuthal asymmetry seen in the \tw\ disk.  The
coronagraphic profiles show that the disk is brighter on one side than
the other between 65 and 140~AU, with the maximum brightness occurring
at PA = $233.^{\circ}6 \pm 5.^{\circ}7$ (including the systematic
uncertainty).
We are not sure exactly how far in toward the star this asymmetry extends, 
due to the limitation of the wedges in the coronagraphic images and 
the possible misalignment in the y-direction of the \tw\ and PSF 
spectroscopic fiducial images.

Forward-scattering can cause a sinusoidal brightness asymmetry between
the near and far sides of an inclined, optically thick disk
(e.g. GG~Tau~A; \citet{Mccabe:2002}).  However, we must explain why the
asymmetry is only seen interior to 140~AU.  The first suggestion is that
the grains in the inner disk are more strongly forward scattering.
However, the disk color spectra show no change between 55 and 155~AU,
above or below the fiducial (see Figure~\ref{fig:color}), which leads us
to believe that the asymmetry is not due to a change in grain
composition or size at 140~AU.  The second suggestion is that the disk
inclination is larger interior to this radius, i.e.\ that the 
disk is warped.  
In this scenario, the direction of maximum brightness  
is the minor axis of the inclined inner disk, 
and the PA of the major axis is $323.^{\circ}6 \pm 5.^{\circ}7$.  
This PA roughly agrees with that deduced from sub-mm CO emission
observations showing the Keplerian rotation of the disk 
($-45^\circ = 315^{\circ}$; \citet{Qi:2004}).

If the asymmetry is due to forward-scattering, then its magnitude 
may be used to calculate a lower limit to the inner disk inclination.
The minimum to maximum surface brightness ratio from the sine
fitting discussed in Section~\ref{sub:coron_results} is
$S_\mathrm{min}/S_\mathrm{max} = 0.52 \pm 0.16$, including the 
systematic error.
Using a Henyey-Greenstein scattering phase function \citep{Henyey:1941}, 
the ratio of the minimum to maximum surface brightness is 
\begin{equation}
S_\mathrm{min}/S_\mathrm{max} \ = \
\frac{[ \, 1 + g^2 - 2 g \cos \, (90^{\circ} - i) \, ]^{3/2}}{%
[ \, 1 + g^2 - 2 g \cos \, (90^{\circ} + i) \, ]^{3/2}} \ ,
\end{equation}
where $g$ is the dust asymmetry parameter ($g = 1.0$ indicates purely
forward-scattering, $g = 0$ indicates isotropic scattering) and $i$ is
the disk inclination ($i = 0^{\circ}$ corresponds to face-on).  
Using the maximum possible value of $g = 1.0$, we find an inner disk 
inclination of $12.^{\circ}4 \, ^{ \: +6.^{\circ}8} _{ \: -5.^{\circ}0}$.  
This is barely consistent with the inclination of $7^{\circ} \pm 1^{\circ}$
found by \citet{Qi:2004} from their sub-mm CO data.  
However, note that the sub-mm data probe a wide range of radii in the disk 
and
give some sort of mean disk inclination.  

Using a more plausible value of $g = 0.5$, we find an inclination of 
$15.^{\circ}6 \, ^{ \: +8.^{\circ}7} _{ \: -6^{\circ}.4}$.  
Our lower limit ($i > 9^{\circ}$) is significantly greater
than the upper limit on the inclination found from isophotal-fitting to
the \tw\ NICMOS imaging data ($i < 4^{\circ}$; \citet{Weinberger:2002}).
Our data have 50\% better spatial resolution than the NICMOS coronagraphic 
images.  
Given that the region over which the disk appears asymmetric is small,
i.e. 65--140 AU or only 1\farcs3, the STIS images are uniquely suited
for measuring the inclination.

A warped inner disk has been observed in the \bpic\ debris
disk (see Figure~8 in \citet{Heap:2000}). Two possible explanations have
been put forward for disk warps \citep{Heap:2000}.  In the first, warps
form through radiation-induced instability in accretion disks
surrounding luminous stars \citep{Armitage:1997}.  The problem with this
explanation is that the star must have a luminosity greater than about
$10 \ L_\sun$, while the \tw\ luminosity is only about 0.25~$L_\sun$
\citep{Webb:1999}.  In the other explanation, the warp is caused by the
gravitational perturbation of a giant planet on an inclined orbit.  The
presence of a protoplanet in the \tw\ disk has already been suggested to
explain the dearth of material within 4~AU of the star
\citep{Calvet:2002}.  But it is not clear if a single protoplanet can
explain both the central hole and the disk azimuthal brightness
asymmetry.

\subsection{Grey Scattering Over the Bulk of the Disk \label{sub:gray}}

The scattering over the bulk of the disk is clearly wavelength-independent
in the optical, in contrast to the results of \citet{Krist:2000}, who 
found the disk to be bluer than the central star.
To determine what this might mean for the properties of the
\tw\ dust, we examine the \citet{Mccabe:2002} Monte Carlo 
scattering simulation of GG~Tau~A, a classical T~Tauri star binary system 
surrounded by a massive disk.

The scattered surface brightness at any one angle, $S_{\mathrm{scatt}}$,
is proportional to $\omega^{n(g_\lambda)}$, where $\omega$ is the grain 
albedo and $n$ is the number of scattering events.
An increase in $n$ produces a decrease in $S_{\mathrm{scatt}}$.
The number of scattering events depends on the dust asymmetry parameter,
$g_\lambda$, which is a function of wavelength.
Note that in the case of \tw, the relatively low inclination of the disk 
indicates that the observed photons have all scattered through almost the 
same angle ($\approx 90^\circ$).

The wavelength-dependence of the scattered light is determined by the
combination of the wavelength-dependences of $\omega$ and $g_\lambda$.
The wavelength-dependence of the albedo tends from blue 
(Rayleigh scattering; small particles) to neutral (large particles).
The dust asymmetry parameter increases with the size parameter
$x = 2 \pi a / \lambda$, reaching a roughly constant value 
at $x \gtrsim 4$ \citep{Wolff:1998}.
Since an increase in $g_\lambda$ produces an increase in $n$ and a
decrease in $S_{\mathrm{scatt}}$, the effect of the dust asymmetry 
parameter is red (small particles) to neutral (large particles) 
scattering.
Therefore, an optically thick scattering situation may produce any color 
light.
Which parameter dominates depends on the number of scattering events.
For example, in a relatively low density case (low $n$), the 
wavelength-dependence of $S_{\mathrm{scatt}}$ is dominated by $\omega$.
The scattered light will be blue to neutral, depending on the particle size. 
In a high density case (high $n$), $g_\lambda$ will dominate, and the 
scattered light will be red to neutral. 

So there are two ways to get neutral scattering from a disk. 
In the first, the number of scattering events is such that a blue albedo is 
exactly canceled by a red dust asymmetry parameter; this implies relatively
small scattering particles.
In the other, the scattering grains are so large that $\omega$ and
$g_\lambda$ are wavelength-independent.
For the first scenario to produce neutral scattering from the \tw\
disk over such a large range of radii ($\approx 100$~AU), the number
of scattering events would have to be just right and not change much  
with radius. 
It seems more likely that the second scenario applies and relatively
large grains are present between about 55 and 155~AU.
Our disk spectra show that the bulk of the disk neutrally scatters 
wavelengths between about 5500 and 1~\um; 
this indicates grains larger than $\frac{x \, \lambda}{2 \pi} \approx
\frac{4 \ (1 \mu\mathrm{m})}{2 \pi} \approx 0.6$~\um.
Since the disk appears to scatter neutrally all the way into the near-IR 
(1.6~\um; \citet{Weinberger:2002}), the particle size is likely greater 
than 1~\um.
These arguments and conclusions will be quantitatively tested in a future
paper by application of the \citet{Mccabe:2002} Monte Carlo scattering
code to the \tw\ disk.

\subsection{Scattering Change in Innermost Region of Disk \label{sub:blue}}

We now consider what the apparent change in the wavelength-dependence of
the scattered light interior to 55~AU  might mean. 
From the above arguments, blue scattering occurs when the number of
scattering events and the grain sizes are both relatively small.
If we imagine for a moment that the the scattering particles are small
over the whole disk, the blue scattering at small radii could be 
caused solely by a decrease in the number of scattering events with 
decreasing radii.
However, since we previously argued that the grains in the outer
regions of the disk are large, the change in scattering at small radii
must also indicate a decrease in particle size.

The idea that there are more small particles close to the star at first seems
counterintuitive, since the coagulation of condensed grains occurs faster 
at smaller radii.
However, the destruction of planetary material also occurs first at smaller
radii, as evidenced by central holes in many debris disks.
Sub-micron grains in debris disks, like the 12~Myr old \bpic\ system 
\citep{Zuckerman:2001a}, must be produced by collisions among 
planetesimals \citep{Artymowicz:1988}.
We therefore suggest that blue scattering at small radii in the \tw\ disk 
may be evidence that the clearing out of solid planetary material has begun 
in the inner disk.

\section{Concluding Remarks} \label{sec:conc}

These observations demonstrate the usefulness of high-contrast 
coronagraphic spectroscopy of scattered light from CS disks.
We here summarize our principal observational results from this study of
the \tw\ protoplanetary disk.

\begin{enumerate}
\item Interior to 140~AU, the disk surface brightness 
shows a sinusoidal dependence on azimuthal angle.
The maximum brightness is at PA = $233.^{\circ}6 \pm 5.^{\circ}7$.
The origin of this feature requires further study, but we  
suggest it is caused by an increase in inclination in the inner part 
of the disk, i.e.\ a warp.
\item The disk spectra definitively show that the scattering is 
wavelength-independent in the optical between 55~AU to 155~AU from the star.
We argue this indicates that the scattering grains are larger than about
1~\um\ out to large radii. 
\item The color of the scattered light from the innermost portion of the
disk (34.5 -- 46.0~AU) appears bluer than the outer disk. 
This result is tentative, due to possible systematic errors in
PSF subtraction close to the fiducial bar.
However, taken at face value, the change in color might indicate a
decrease in grain size at small radii.
\end{enumerate}

\begin{figure}
\plotone{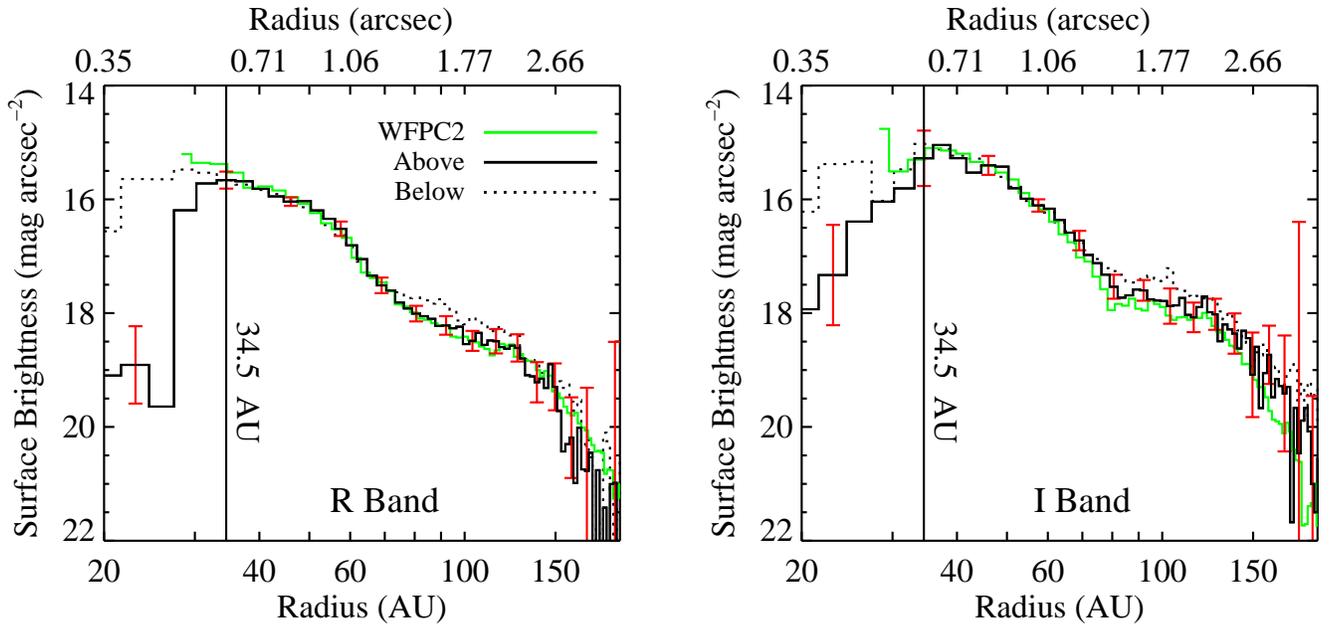}
\caption{R and I band radial surface brightness profiles of the \tw\ disk.
The profiles above the fiducial are shown with black solid lines, the 
profiles below with black dashed lines.
The error bars (in red) are the total errors, including systematic errors.
The \hst\ WFPC2 profiles are overplotted with green solid lines.
These are the profiles shown in Figure~5 of \citet{Krist:2000}, converted from
WFPC2 F606W and F814W band fluxes per arcsec$^{2}$ to R and I band magnitudes
per arcsec$^{2}$.
No arbitrary shifting or scaling was applied to the WFPC2 profiles.
\label{fig:RI}}
\end{figure}

\begin{figure}
\plotone{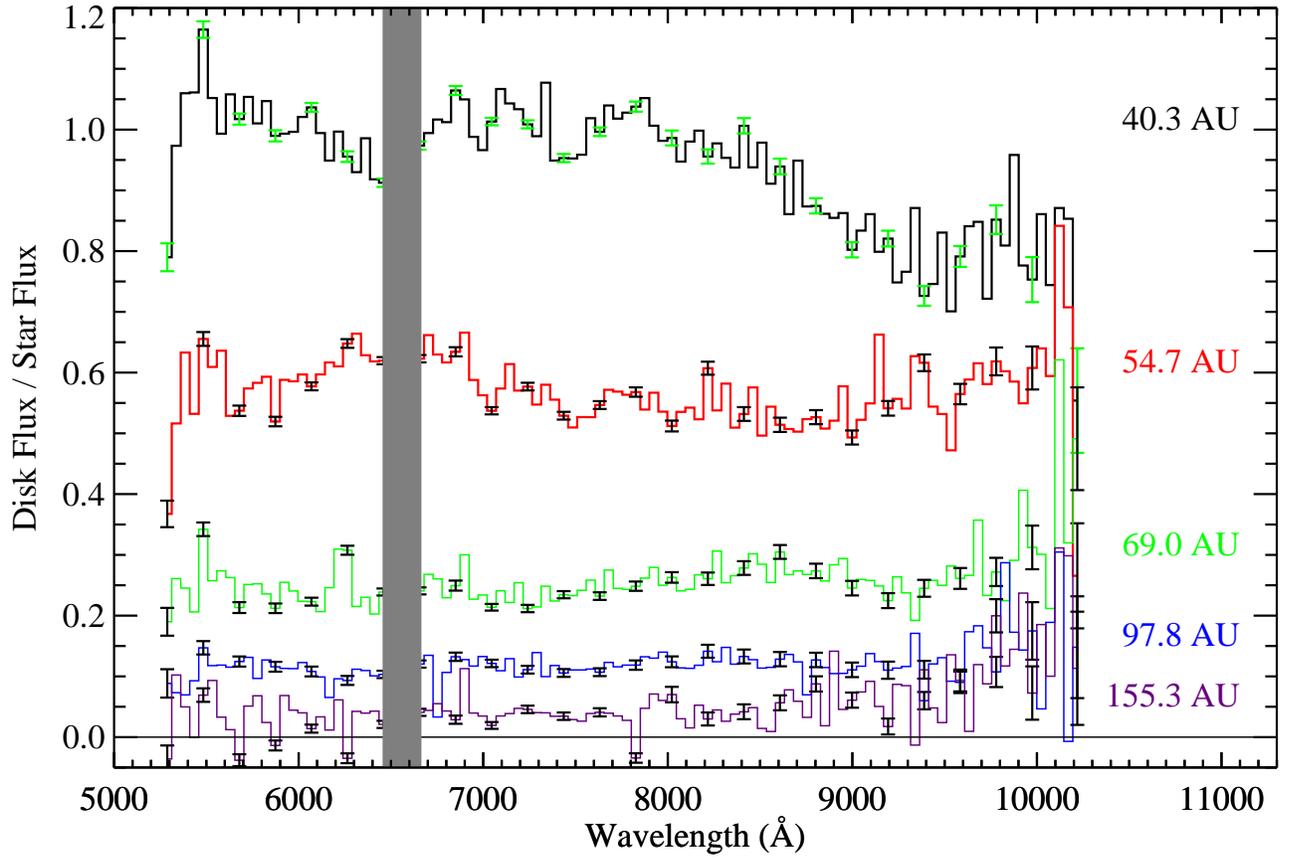}
\caption{Ratio of \tw\ disk spectra to stellar spectrum at increasing radii
(referred to in the text as the disk color spectra). 
The y-axis is the extracted disk flux divided by 
$2.95 \times 10^{-4}$ times the \tw\ point-source spectrum. 
The data have been rebinned by a factor of 10 to increase the signal-to-noise.
The region around \ha\ has been blocked out with a gray bar 
(see Section~\ref{sub:psf}).
The error bars are the $\pm 1 \sigma$ propagated statistical errors.
\label{fig:color}}
\end{figure}

\begin{figure}
\plotone{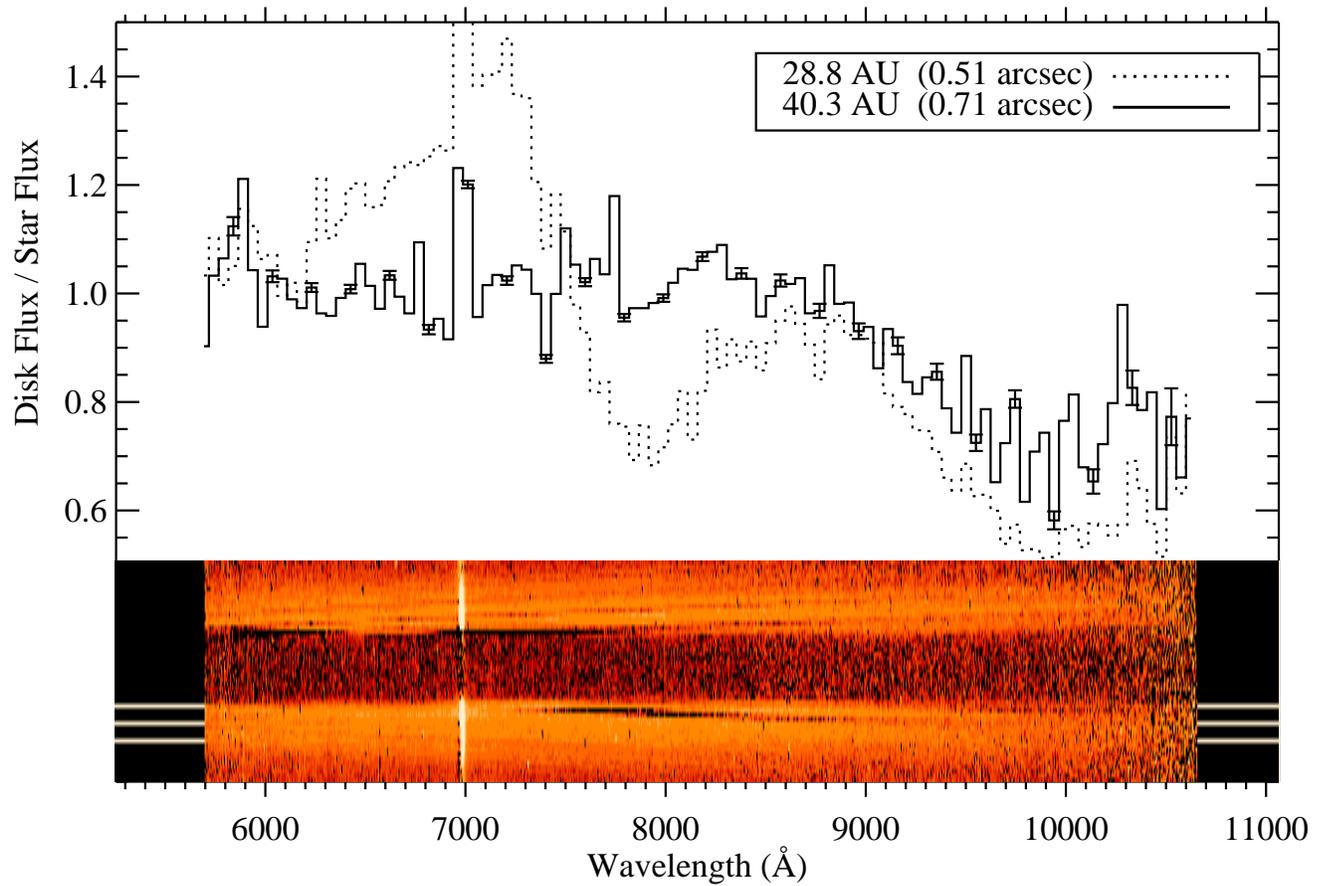}
\caption{Effect of the worst PSF-subtraction residual on the extracted 
disk color
spectra.
The PSF-subtracted 2-D disk spectrum is shown at the bottom, with
the same brightness scaling as in Figure~\ref{fig:2d}.
The spectra extracted at 28.8~AU (dashed line) and 40.3~AU (solid line)
are shown at the top, on the same wavelength scale as the 2-D spectrum.
The white bars at the edges of the 2-D spectrum show the limits of the
extraction boxes in the y-direction.
The worst PSF-subtraction residual is visible in the 2-D spectrum as
a dark stripe below the fiducial bar; its effect may be seen in the
28.8~AU spectrum as a dip between about 7000~\AA\ and 8400~\AA.
\label{fig:artifact}}
\end{figure}

\begin{figure}
\epsscale{0.8}
\plotone{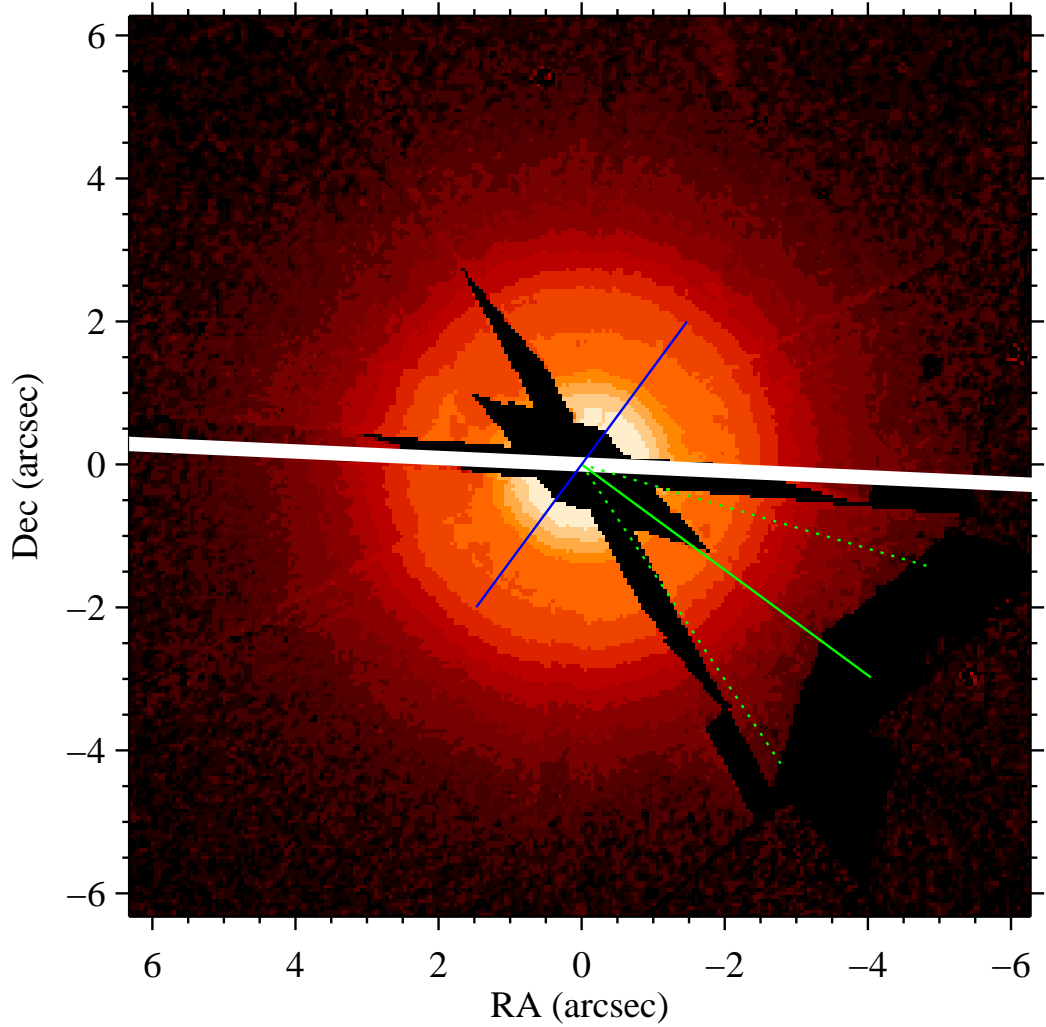}
\caption{STIS CCD coronagraphic image of \twhya.
The image shows a logarithmic scaling of arbitrary surface brightness units.
Locations where there are no data due to the coronagraphic wedge or 
the diffraction spikes are masked out in black.
The $0\farcs2$ wide slit oriented at a position angle of $87.^{\circ}4$ E~of~N
used in the spectroscopic observations is shown in white.
The direction of maximum disk brightness (PA $= 233.^{\circ}6$) 
found from the sine fitting in Figure~\ref{fig:best_asym} is 
indicated with a solid green line of length $5\farcs02 = 283$~AU;
this is the outermost radius at which the disk is detected
(see Figure~\ref{fig:coron_prof}).
The dashed green lines show PA $= 233.^{\circ}6 \pm 20^{\circ}$;
pixels between these position angles were included in the maximum
brightness radial profile shown in Figure~\ref{fig:coron_prof}.
Our suggested major axis of the inclined inner disk is shown with a
solid blue line of length $2\farcs48 = 140$~AU;
this is the outer edge the inclined inner disk.
\label{fig:coron}}
\end{figure}

\begin{figure}
\epsscale{1.0}
\plotone{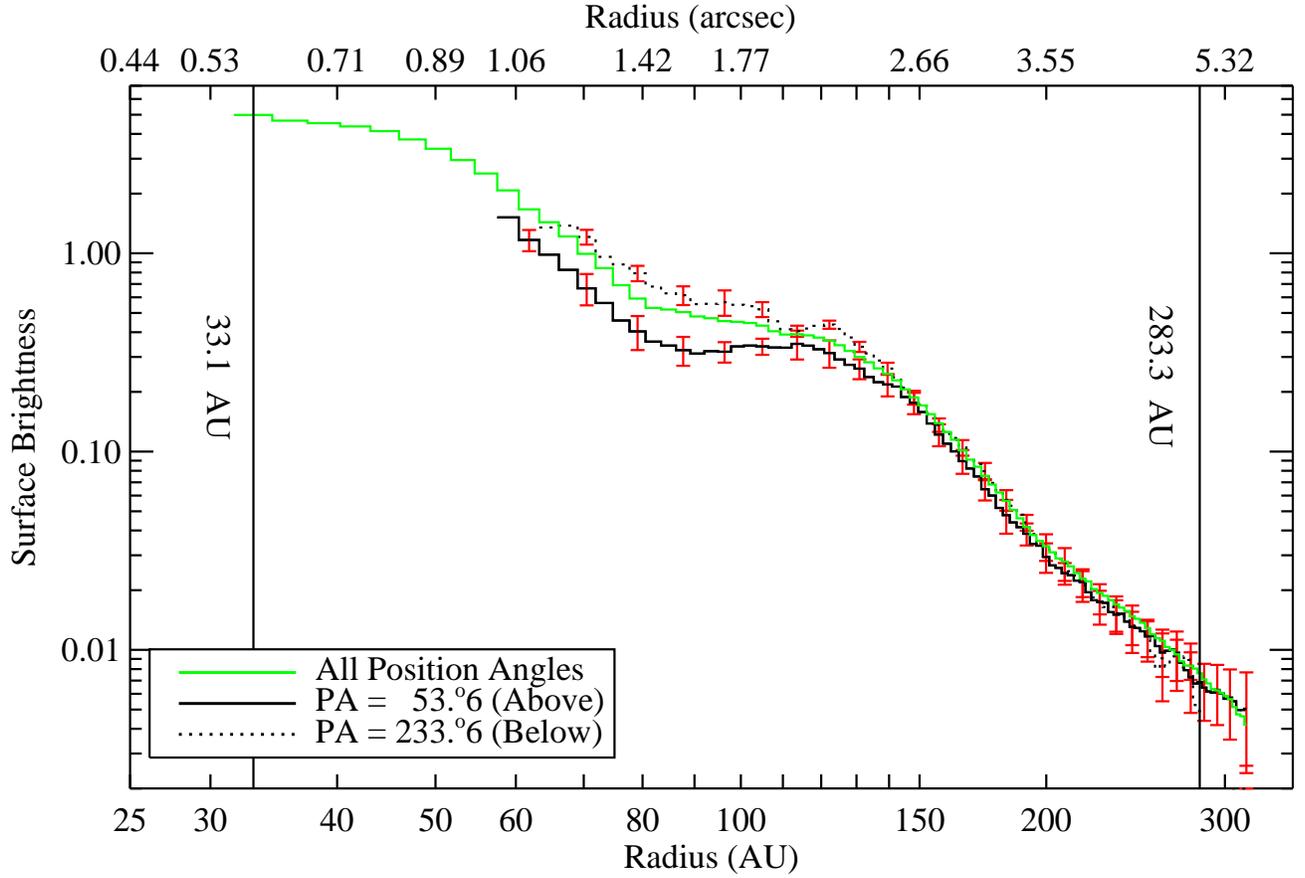}
\caption{Radial surface brightness profiles from the STIS coronagraphic image
of \twhya\ shown in Figure~\ref{fig:coron}.
Surface brightness in arbitrary units appear on the y-axis, and distance
from the star on the x-axis.
The azimuthally-averaged profile using all pixels 
available at each radius is shown with a green line.
The disk is detected between 33.1 and 283.3~AU ($3 \sigma$ cutoff).
The solid and dashed black lines show the profiles which include
only pixels with position angles within $\pm 20^{\circ}$ of
the directions of maximum and minimum brightness found from the sine 
fitting in Figure~\ref{fig:best_asym}.
The $\pm 1 \sigma$ statistical error bars on these two curves are shown in red.
The position angle of the spectroscopic slit ``below the fiducial'' 
(PA $= 267.^{\circ}4$) is close to the direction of maximum brightness, and
the radial profiles from the spectroscopic data shown in 
Figure~\ref{fig:tot_prof} are qualitatively similar to those shown here.
\label{fig:coron_prof}}
\end{figure}

\begin{figure}
\epsscale{0.8}
\plotone{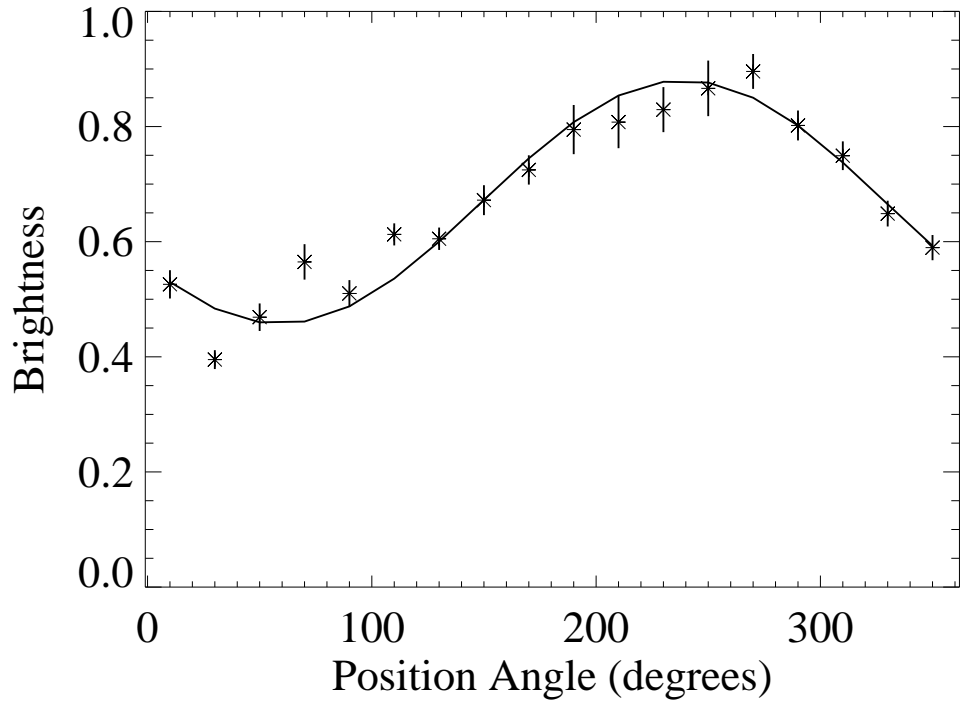}
\caption{Brightness of the \twhya\ disk in the STIS coronagraphic
image shown in Figure~\ref{fig:coron} as a function of position angle.
The points show the mean disk brightness between 70 and 88~AU 
in $20^{\circ}$ intervals;
the error bars are the $\pm 1 \sigma$ statistical errors.
The solid line shows the weighted least-squares sine fit to the data.
The directions of maximum and minimum brightness are 
$233.^{\circ}6 \pm 5.^{\circ}7$
(including systematic errors)
and $53.^{\circ}6 \pm 5.^{\circ}7$ East of North.
\label{fig:best_asym} }
\end{figure}

\acknowledgments
We thank John Krist for providing us with the WFPC2 radial surface 
brightness profiles of \tw\ for inclusion in Figure~\ref{fig:RI}.
This paper is based on observations made with the NASA/ESA 
\emph{Hubble Space Telescope}, obtained at the Space Telescope Science 
Institute, which is operated by the Association of Universities for 
Research in Astronomy, Inc., under NASA contract NAS 5-26555. 
These observations are associated with program GO-8624;
we thank our co-investigators on this program for useful discussions.

%%
%% REFERENCES
%% 

%%
%% TABLES
%%

\begin{deluxetable}{lccclc}
\tablecaption{Observation Log \label{tab:log}}
\tablewidth{0pt}
\tablecolumns{6}
\tablehead{
\colhead{ID} & \colhead{Target}    & \colhead{Time} & \colhead{Exp. Time} 
& \colhead{Aperture} & \colhead{Comment} \\
\colhead{ } & \colhead{ } & \colhead{(UT)} & \colhead{(s)} & \colhead{ } & \colhead{ } 
}
\startdata
\multicolumn{6}{c}{Spectroscopy} \\
\hline
\sidehead{2002 July 17}
o64w57010 & \tw		& 02:48:46	& 72 	& 52 $\times$ 0.2   & point source \\
o64w57020 & Lamp Flat	& 02:52:08	& 25	& 52 $\times$ 0.2   & \\
o64w57030 & \tw		& 04:03:32	& 198	& 52 $\times$ 0.2F2  & fiducial \\
o64w57040 & ''		& 04:09:02	& 1380	& 52 $\times$ 0.2F2  & '' \\
o64w57050 & ''		& 05:44:00	& 1380	& 52 $\times$ 0.2F2  & '' \\
o64w57060 & ''		& 06:13:18	& 1221	& 52 $\times$ 0.2F2  & '' \\
o64w57070 & ''		& 07:25:00	& 1380	& 52 $\times$ 0.2F2  & '' \\
o64w58010 & CD-43 2742	& 00:38:55	& 36	& 52 $\times$ 0.2   
& PSF star, point source  \\
o64w58020 & Lamp Flat	& 00:41:41	& 25	& 52 $\times$ 0.2   &   \\
o64w58030 & CD-43 2742	& 00:44:41	& 684	& 52 $\times$ 0.2F2 
& PSF star, fiducial \\
o64w58040 & ''		& 01:00:29	& 665	& 52 $\times$ 0.2F2 & '' \\
\hline
\multicolumn{6}{c}{Imaging} \\
\hline
\sidehead{2000 November 3} 
o64w04010 & \tw         & 21:01:21      & 904   & WEDGEA1.0         & \\
o64w04020 & ''          & 21:19:21      &1260.2 & WEDGEA1.0         & \\
o64w05010 & HD 85512    & 22:29:20      & 416   & WEDGEA1.0         & PSF star\\
o64w05020 & ''          & 22:42:08      & 416   & WEDGEA1.0         & '' \\
o64w05030 & ''          & 22:54:56      & 416   & WEDGEA1.0         & '' \\
o64w05040 & ''          & 23:07:44      & 156   & WEDGEA1.0         & '' \\
\sidehead{2000 December 27}
o64w06010 & \tw         & 13:51:48      & 904   & WEDGEA1.0         &\\
o64w06010 & \tw         & 14:09:48      &1260.2 & WEDGEA1.0         &\\
\enddata
\tablecomments{All spectroscopic data were taken with the STIS CCD at
GAIN=1, using the G750L grating. All imaging data were taken with the 
unfiltered
(clear) CCD at GAIN=4.}
\end{deluxetable}
\clearpage

%%
%% FIGURES
%%
%% Use the figure environment and \epsfig or \plottwo to include 
%% figures and captions in your electronic submission.

\end{document}